\documentclass[journal=jpcl,manuscript=letter]{achemso}

\usepackage[version=3]{mhchem} 

\author{Jeongmin Kim}
\email{jeongmin@pusan.ac.kr}
\affiliation[pusan]
{Department of Chemistry Education and Graduate Department of Chemical Materials, Pusan National University, Busan, 46241, Republic of Korea}
\author{Bong June Sung}
\email{bjsung@sogang.ac.kr}
\affiliation[sogang]
{Department of Chemistry and Institute of Biological Interfaces, Sogang University, Seoul 04107, Republic of Korea}
\date{\today}

\title[]{Subcritical Pitchfork Bifurcation Transition of a Single Nanoparticle in Strong Confinement}

\abbreviations{IR,NMR,UV}
\keywords{American Chemical Society, \LaTeX}

\begin{document}

\begin{tocentry}
\begin{center}
\includegraphics[width=0.65\linewidth] {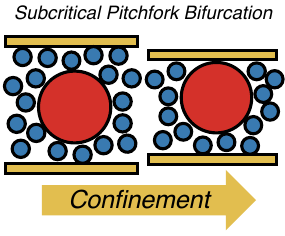}
\label{fig:toc}
\end{center}



\end{tocentry}

\begin{abstract} 
Confinement influences fluid properties. We show, employing molecular dynamics simulations with explicit solvents, that slit confinement drives a first-order transition for a small nanoparticle between staying at the slit center and binding to the slit surfaces. The transition follows a subcritical pitchfork bifurcation, accompanying a similar transition of the nanoparticle's lateral diffusion, depending on interparticle interactions and confinement interfaces. Our findings underscore the necessity for advancing molecular hydrodynamics under strong confinement.
\end{abstract}

Confined systems are ubiquitous including lab-on-a-chip devices~\cite{hopkins2021tristability,haward2021bifurcations,haward2021stagnation}, nanofluidic devices~\cite{bocquet2010nanofluidics,bocquet2020nanofluidics}, porous media~\cite{bousige2021bridging}, optical trap~\cite{metzger2006observation,franosch2011resonances,bustamante2021optical}, and biological capillaries~\cite{WhitesidesNature,SquiresRMP,SchochRMP,HoltScience,BurgNature,HaywoodAnalChem,HaScience,LevingerScience,BerndsenPNAS,FerraroPRL}. When fluids are subjected to confinement, their properties undergo significant changes~\cite{liquids,KogaPRL,RodriguezPRL,ChoudhuryJCP,MithemJCP,DyrePRL2013,LengPRL,MittalJCP,MittalPRL,RichertAnnuRev,PeylaEPL,PeylaEPL83,kavokine2021fluids}. For example, in the blood vessel, blood viscosity exhibits a singularity as the distribution of red blood cells changes with their volume fraction~\cite{MisbahPRL}. Thin-film confinement also significantly affects the mechanical and electrical properties of polymer films, critically depending on the dispersion of nanoparticles~\cite{OchoaJPCM,GoswamiPRE,XueProgress,GuoPRL,KrishnanJPCM,WangSoftMatter,OpfermanPRE,TitekaPRL,BasuSoftMatter2015,ChandranNatComm2014,KrishnanScience2006}. With recent developments in nanofluidic experiments, theoretical understanding is under active development with growing interest, particularly bridging molecular foundations and continuum-level descriptions~\cite{lesnicki2016molecular,bocquet2010nanofluidics,straube2020rapid,kavokine2021fluids,aluru2023fluids,kellouai2024gennes}.

In general, equilibrium states (stable fixed points) and their transitions of nonlinear systems (including confined fluids) can be determined using a nonlinear differential equation of a reaction coordinate of interest~\cite{sole2011phase,barrat2003basic}. Moreover, a nonlinear dynamics provides a unified framework to understand rich phenomena of seemingly unrelated systems. For instance, a subcritical pitchfork bifurcation (SPB, Eq.~\ref{eq:spb})~\cite{colet1990theory} is an excellent model for understanding phase transitions across diverse systems 
from the gas-liquid transition~\cite{barrat2003basic} to the on-resonance single-mode laser with a saturable absorber~\cite{lsa78,colet1990theory} and the gene expression of the clonal cell population~\cite{development09}.
\begin{eqnarray}\label{eq:spb} 
\frac{dz}{dt}=\frac{dF(z)}{dz}= -\alpha_1(H)z + \alpha_3z^3- \alpha_5z^5.
\end{eqnarray}
Here, $z$ is the reaction coordinate, $\alpha_3$ and $\alpha_5$ are positive constants, and $\alpha_1$ is a constant, whose value depends on an external control parameter $H$. The fixed points $z^*$ are the states of a potential $F(z)$ of mean force, satisfying $\frac{dz}{dt}|_{z=z^*}=\frac{dF(z)}{dz}|_{z=z^*}=0$. The number and stability of fixed points depend on the constants and can change across the phase transitions at bifurcation points. As one modulates $H$, the number of stable and metastable fixed points changes from one to three, like a pitchfork. When $\alpha_1(H)>0$, the fixed point $z^*=0$ is stable, while metastable when $\alpha_1(H)<0$. For instance, with $\alpha_1=a(\frac{1}{H}-\frac{1}{H_0})$, $\alpha_3=b$ and $\alpha_5=1$, Eq.~\ref{eq:spb} of the SPB has three bifurcation points ($H^*>H_c>H_0$) that satisfy ${H_0}^{-1}-{H_c}^{-1}=3b^2/16a$ and ${H_0}^{-1}-{H^*}^{-1}=b^2/4$.

In this Letter, we show that strong slit confinement can drive a first-order transition across a critical slit size $H=H_c$ in the spatial distribution of a small nanoparticle, normal to the surfaces. Such a discontinuous transition qualitatively influences both the solvation state and the lateral diffusion of the nanoparticle. Our systematic investigation reveals that the phase transitions driven by a slit size $H$ well follow the SPB transition (Eq.~\ref{eq:spb}) under various conditions, indicating such a transition belongs to the same universality class with \emph{e.g.}, a gas-liquid first-order phase transition.

The near-wall diffusion of a Brownian particle differs from its bulk diffusion, being hindered, position-dependent, and even Fickian-yet-non-Gaussian~\cite{BoonYip,liquids,BatchelorJFMech,JRJCP,huang2015effect,mo2019highly,alexandre2023non}. The interface, suppressing the long-wavelength hydrodynamic modes, slows down the lateral diffusion of the nanoparticle than at the bulk~\cite{bocquet1995diffusive,chio2020hindered}: the lateral diffusion coefficient ($D_\parallel$) of a Brownian particle near a wall decreases monotonically as the distance from the wall decreases. The hindered near-wall diffusion was also captured by a faster algebraic decay of the velocity autocorrelation function ($C_v(t)\sim t^{-f}$) in a long-time limit with $f=5/2$ than the bulk diffusion with $f=3/2$~\cite{pagonabarraga1998algebraic,huang2015effect}. In this work, we find a dynamic transition of a small nanoparticle in strong confinement (\emph{e.g.}, breakdown of the linear superposition approximation~\cite{chio2020hindered} due to significant hydrodynamic interactions with both walls), evident by the non-monotonic trend of $D_\parallel$ with the confinement gap $H$. This exotic dynamic transition occurs due to the SPB transition of the spatial distribution of the nanoparticle, which accompanies the change in its solvation state.

\begin {figure*}
\includegraphics [width=1\linewidth] {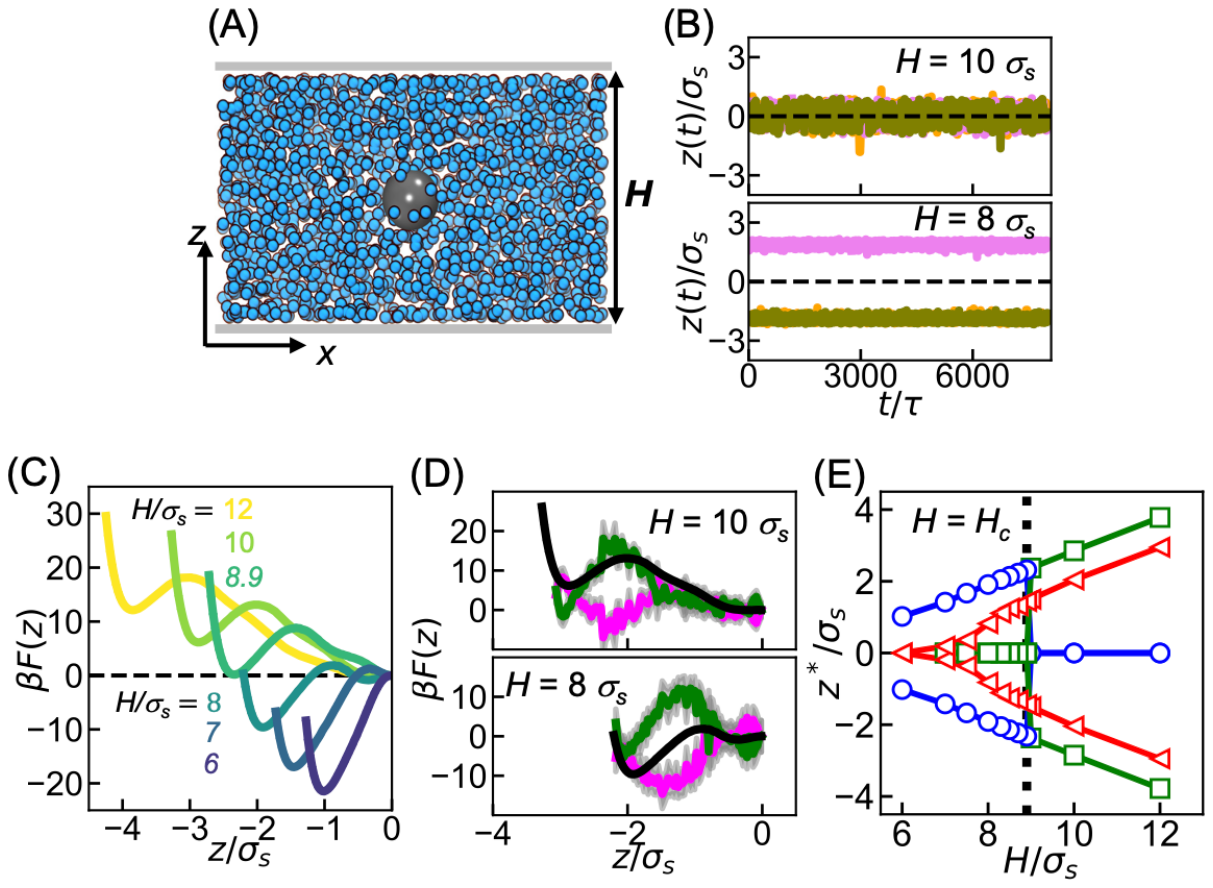}
\caption{Subcritical pitchfork bifurcation of a nanoparticle in a smooth slit of a gap $H$. 
Here, $r_c = 2.5\sigma_{ns}$ (see Eq.~\ref{eq:lj}).
(A) A schematic of the confined nanoparticle solution. (B) The vertical position $z(t)$ of the nanoparticle 
for $H=10~\sigma_s$ and $8~\sigma_s$. The black dashed lines indicate the slit center at $z=0$. Different colors represent different trajectories. (C) Potential $F(z)$ of the mean force 
(D) Energetic ($\beta U(z)$, pink) and entropic ($-S(z)/k_B$, green) contributions to $F(z)~(\equiv U(z)-TS(z))$ for $H=10$ and $8~\sigma_s$,
relative to their value at $z=0$. The grey-shaded areas represent the statistical error with the standard deviation. (E) The bifurcation diagram obtained from $\beta F(z)$ in panel (C).
Blue, green, and red markers indicate stable, metastable, and unstable fixed points, respectively. 
The vertical dotted line represents the critical gap $H=H_c$, across which the first-order bifurcation occurs; see the main text for the details.
}
\label{fig:bifurcation}
\end{figure*}

We consider the system that consists of a single nanoparticle and solvent particles confined between two parallel walls (Fig.~\ref{fig:bifurcation}A). 
Interactions between particles are described by the standard truncated and shifted Lennard-Jones (LJ) potential ($U_{ij}$)~\cite{FrenkelSmit}. 
\begin{equation}\label{eq:lj}
U_{ij}(r_{ij}) = 4\epsilon_{ij} \left[\left( \frac{\sigma_{ij}}{r_{ij}} \right)^{\alpha_r} - \left({\frac{\sigma_{ij}}{r_{ij}}}\right)^{\alpha_a}\right] - U_c ~\text{if}~ r_{ij} < r_c, 
\end{equation}
where $\alpha_r$ = 12 and $\alpha_a$ = 6. $U_c = 4\epsilon_{ij} [(\frac{\sigma_{ij}}{r_{c}})^{\alpha_r} - {(\frac{\sigma_{ij}}{r_{c}}})^{\alpha_a}]$ and $r_c$ denotes the cut-off distance such that $U_{ij}(r_{ij}\geq r_c)=0$. The indices $i$ and $j$ represent either the nanoparticle ($n$) or a solvent particle ($s$). The mass $m_s$ and the diameter $\sigma_s$ of solvent particles are the units of mass and length, respectively.
The nanoparticle is chosen to be 5 times as large ($\sigma_n=5\sigma_{s}$) and 125 times as heavy ($m_n=125m_s$) as a solvent particle, so it can be considered a Brownian particle in the bulk solution, satisfying the Stokes-Einstein relation~\cite{LeeTheoChem,JRJCP}.
$k_BT~(\equiv\beta^{-1})$ is the unit of energy with $k_B$ the Boltzmann constant and $T$ temperature. The unit ($\tau$) of time is, then, $\tau\equiv\sigma_{s}\sqrt{m_s/k_BT}$. We tune the interaction ($U_{ns}$) between the nanoparticle and solvent particles by changing the values of $r_c$, while keeping $\sigma_{ns}=3\sigma_{s}$. In case of $r_c=2.5\sigma_{ns}$, $U_{ns}$ is attractive at intermediate distances, while $U_{ns}$ becomes purely repulsive with $r_c=1.122\sigma_{ns}$. 
Unless otherwise noted, $\epsilon_{ij}=k_BT$.  

We place two parallel walls in $xy$ planes at $z = \pm\frac{H+\sigma_s}{2}$ to construct the slit confinement: their gap $H$ between two walls, ranges from 6 to 16$\sigma_{s}$. This study considers two types of walls: smooth and corrugated. In the case of smooth walls, the interaction between the wall and a particle is also described by Eq.~\ref{eq:lj} but with $\alpha_r=10$ and  $\alpha_a=4$~\cite{GelbJCP}. We construct two kinds of corrugated walls by placing and fixing spherical wall particles in a single layer of square or hexagonal lattice sites with the same number ($N_w$) of the wall particles: the total number of wall particles is $2N_w=512$. The diameter $\sigma_w$ and the mass of the wall particle are equal to those of the solvent particles. Interactions between wall particles ($w$) and other particles (either the nanoparticle $U_{nw}$ or the solvent $U_{sw}$) are also described by $U_{ij}$ with $\alpha_r=12$, $\alpha_a=6$, and $\epsilon_{nw}=\epsilon_{sw}=k_BT$. Unless otherwise noted, $U_{nw}$ and $U_{sw}$ are purely repulsive with a cut-off distance $1.122\sigma_{nw}$ and $1.122\sigma_{sw}$, respectively.

The nanoparticle is placed initially at the slit center, while solvent particles are distributed randomly in the slit. While tuning $H$, the number $N$ of the solvent particles changes in such a way as to keep their density $\rho=\frac{N}{L^2H}$ constant with $L$ the lateral dimension of the slit. Unless otherwise noted, $\rho=0.74\sigma_s^{-3}$ in this study. The initial momenta of particles are sampled via the Maxwell-Boltzmann velocity distribution with zero total momentum according to the chosen temperature. 

All simulations are conducted under canonical ensemble at temperature $k_BT=1$ using the Large-scale Atomic/Molecular Massively Parallel Simulator (LAMMPS) package~\cite{plimpton1995fast}. We employ the velocity Verlet integrator with a time step of 0.005$\tau$ or 0.01$\tau$. We apply the Nos\'e-Hoover thermostat, relaxing the temperature every 50 integration steps. Periodic boundary conditions are applied in the $x$ and $y$ directions. We compute the potential $F(z)$ of mean force of the nanoparticle by performing umbrella sampling with an additional biasing harmonic potential and the weighted histogram analysis method~\cite{wham}. We take the $z$-position of the nanoparticle (in the direction perpendicular to the walls) as the reaction coordinate.

Figure~\ref{fig:bifurcation} describes our main finding that the nanoparticle undergoes the first-order phase transition in its spatial arrangement across the critical gap value $H_c$. In Figure~\ref{fig:bifurcation}, we consider a nanoparticle with $r_c = 2.5\sigma_{ns}$ confined between smooth walls. As depicted in Fig.~\ref{fig:bifurcation}B, the nanoparticle stays at the center ($z=0$) in a large slit ($H=10\sigma_s>H_c$, upper panel) while near the confinement interface ($z\approx\pm2\sigma_s$) in a narrow slit ($H=8\sigma_s$, lower panel). When $H=8\sigma_s<H_c$, the nanoparticle carries out hopping motions between two stable points near the confinement interfaces if given sufficient time during the simulation.

Figure~\ref{fig:bifurcation}C depicts the potentials $F(z)$ of mean force, revealing $H_c\approx8.9\sigma_s$. $F(z)$ is computed and drawn in the half range ($z<0$) of the confinement due to its parity symmetry. For $H=12\sigma_s>H_c$, the global minimum of $F(z)$ is placed at $z=0$ with two metastable points near the walls. With decreasing $H$, the difference in $F(z)$ between the global and metasable minima decreases. At the critical value of $H=H_c=8.9\sigma_s$, two states (near the walls and at the center) are equally likely. Below $H_c$, the stability of the fixed points is reversed: $z^*=0$ becomes metastable, and the fixed points near the walls become globally stable. Below $H\approx6\sigma$, $z^*=0$ becomes unstable with two global minima near the walls. A similar transition was observed in a recent simulation, which suggested that the solvent-induced interactions could induce kinetic trapping of a nanoparticle near walls, depending on parameters such as Hamaker constants, interfacial free energies, and nanoparticle size~\cite{singletary2024kinetic}.

We find that the computed bifurcation diagram (Fig.~\ref{fig:bifurcation}E) follows the SPB transition described by Eq.~\ref{eq:spb} with, for instance, $\alpha_1=a(\frac{1}{H}-\frac{1}{H_c})$, $\alpha_3=b$, and $\alpha_5=1$. Two features of $F(z)$ are again apparent: (i) the phase transition across $H=H_c\approx8.9\sigma_s$ is first order, \emph{i.e.}, a discontinuous jump in the value of the stable fixed point, and (ii) there are no metastable points at $H=6\sigma_s$, implying that $H_0\approx6\sigma_s$. This agreement with the SPB suggests that the spatial distribution of a nanoparticle in slit confinement is a problem of the same class with \emph{e.g.} the gas-liquid phase transition~\cite{barrat2003basic}.

In order to determine a driving force of the bifurcation across $H_c$, we decompose $F(z)$ into the energetic [$U(z)$] and the entropic [$-TS(z)$] contributions (pink and green lines in Fig.~\ref{fig:bifurcation}D, respectively), where $U(z)$ is the total system energy with the nanoparticle located at $z$ and $-TS(z)$ is computed using $-TS(z)\equiv F(z)-U(z)$. All three quantities ($F(z)$, $U(z)$, and $TS(z)$) are relative to their value at $z=0$. Regardless of $H>H_0$, there is a high entropic barrier of about 10$k_BT$ between the states at the center and the slit wall. On the other hand, $U(z)$ changes qualitatively with $H$: the nanoparticle is energetically stable at the center for large $H$, while it is energetically more stable near the walls for small $H$ (see Supporting information for other $H$'s). This suggests that the SPB transition in our system should be an energy-driven process with an entropic barrier. 

\begin {figure}
\centering\includegraphics [width=1\linewidth] {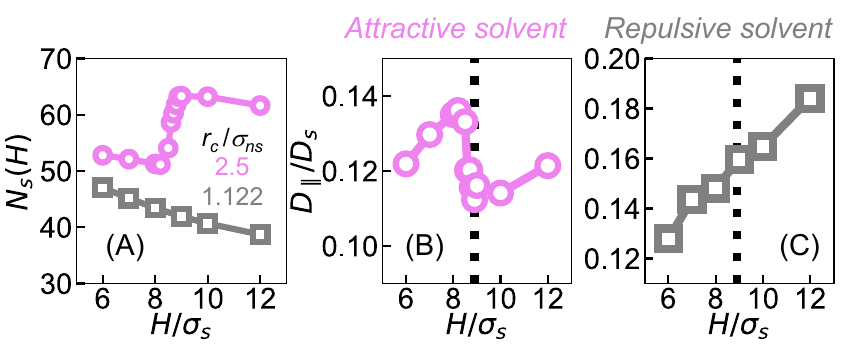}
\caption{Effects of the nanoparticle-solvent interactions $U_{ns}(r)$ on the solvation state and lateral diffusion of the nanoparticle: attractive ($r_c=2.5\sigma_{ns}$, violet circles) and purely repulsive ($r_c=1.122\sigma_{ns}$, grey squares) $U_{ns}(r)$. (A) The number $N_s$ of solvent particles around the nanoparticle within its first solvation layer. (B) and (C) The relative diffusion coefficient ($D_\parallel/D_s$) of the nanoparticle to that of solvent particles.}
\label{fig:diffusion}
\end{figure}

The SPB transition accompanies the transition in the solvation state of the nanoparticle. Fig.~\ref{fig:diffusion}A depicts the number $N_s$ of the solvent particles within the first solvation shell around the nanoparticle as a function of $H$. At $H\approx H_c$, $N_s$ drops significantly with decreasing $H$ (pink symbols in Fig.~\ref{fig:diffusion}A), suggesting that the SPB transition alters the solvation state of nanoparticle across $H=H_c$: Above $H_c$ the nanoparticle residing at $z=0$ is solvated fully by solvent particles while below $H_c$ partially desolvated near the confinement interface. 

Furthermore, we find that the SPB transition dramatically affects the lateral diffusion of the nanoparticle. Fig.~\ref{fig:diffusion}B shows the lateral diffusion coefficient $D_\parallel$ of the nanoparticle, computed from the lateral mean-square displacement, \emph{i.e.}, $D_\parallel\equiv\lim_{t \rightarrow \infty}D_\parallel(t)=\lim_{t\rightarrow\infty}\langle (\vec{r}_{xy}(t) - \vec{r}_{xy}(t=0))^2\rangle/4t$, where $\vec{r}_{xy}(t)$ is the position vector of the nanoparticle in the $xy$ plane at time $t$, and $\langle\cdots\rangle$ denotes the ensemble average. In all cases, the time-dependent lateral diffusion coefficient $D_\parallel(t)$ converges well to its long time limit $D_\parallel$ in our simulation times (see the Supporting information). Similarly, we computed the lateral diffusion coefficient $D_s$ of the solvent.

It is evident that a dynamic transition of the nanoparticle also occurs across $H=H_c$, demonstrated by the discontinuous jump of $D_\parallel/D_s$ at $H\approx H_c$ (Fig.~\ref{fig:diffusion}B). As the ratio $D_\parallel/D_s$ approximates the Stokes-Einstein relation (SER) ($D_s$ is inversely proportional to the solvent viscosity), $D_\parallel/D_s$ is expected to stay constant in bulk solutions. When $H>H_c$, $D_\parallel/D_s\approx 0.12$ regardless of $H$. However, when the SPB transition occurs at $H = H_c$, both $D_\parallel$ and $D_\parallel/D_s$ increase discontinuously. We note that $D_\parallel$ itself also shows a similar jump across $H=H_c$, while $D_s$ decreases monotonically with decreasing $H$ (see the Supporting information).

\begin {figure}
\centering\includegraphics [width=1\linewidth] {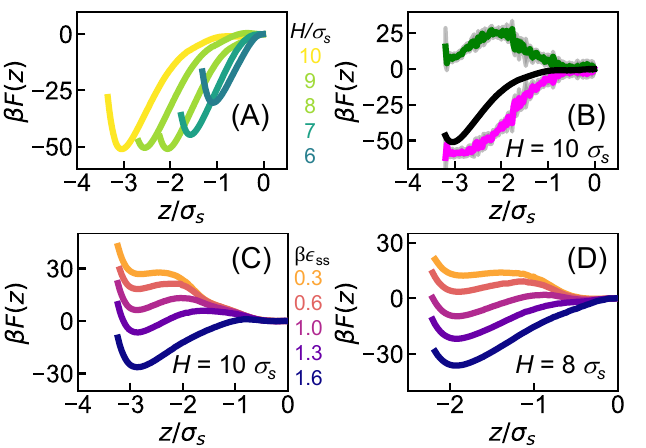}
\caption{Potential $F(z)$ of the mean force for the nanoparticle with various types of solvent. (A) and (B) Purely repulsive $U_{ns}(r)$ with $r_c=1.122\sigma_{ns}$. In panel (B), pink and green solid lines represent the energetic and entropic contributions to $F(z)$ (black solid line), respectively. The grey-shaded areas represent the statistical error with the standard deviation. (C) and (D) Various solvent of different strength $\epsilon_{ss}$ of the solvent-solvent interactions for $H=10~\sigma_s$ and 8~$\sigma_s$.}
\label{fig:interaction}
\end{figure}

\begin {figure}
\centering\includegraphics [width=1\linewidth] {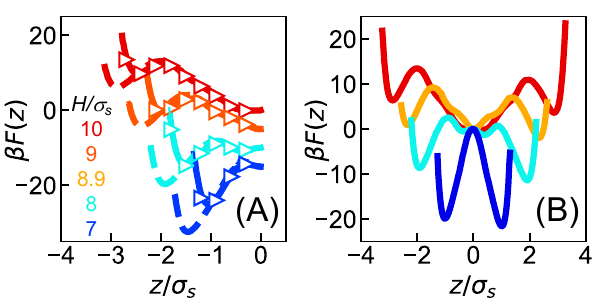}
\caption{
Potential $ F(z)$ of the mean force for the nanoparticle in slit confinement with various interfaces: (A) Different types of interface structures, and (B) asymmetric interfaces. In panel (A), solid lines and square markers represent the hexagonal and square structure of the corrugated walls, respectively, while dashed lines represent the smooth walls. For clarity the curves are displaced vertically from each other. In panel (B), the \emph{left} walls at $z<0$ are purely repulsive, yet the \emph{right} walls at $z>0$ are attractive to the nanoparticle.}
\label{fig:wall}
\end{figure}

For further systematic investigation, we examine two other parameters that alter the confined nanoparticle solutions: the nanoparticle-solvent and solvent-solvent interactions (Fig.~\ref{fig:interaction}) and the confinement interface (Fig.~\ref{fig:wall}). As shall be shown, the SPB well describes the spatial distribution of the nanoparticle in all cases studied in this work. In other words, Eq.~\ref{eq:spb} is applicable for the solutions in other conditions with different $\alpha$ parameters. Our results imply that one may control the SPB transition by modifying the nanoparticle's surface or changing the solvent.

Figure~\ref{fig:interaction} illustrates how the nanoparticle-solvent $U_{ns}$ and solvent-solvent $U_{ss}$ interactions significantly affect the SPB transition. According to the $F(z)$ (Fig.~\ref{fig:interaction}A), when $U_{ns}$ is purely repulsive with $r_c= 1.122\sigma_{ns}$, the nanoparticle stays only near the walls at all $H$'s investigated. No SPB bifurcation with $H$, only with two stable minima near the confinement interfaces, implies $H_0\gg 10\sigma_s$.
As shown in Fig.~\ref{fig:interaction}B, the sizeable energetic gain (magenta line) drives the nanoparticle to be near the walls for $H=10\sigma_s$; the relatively attractive $U_{ss}$ dominates the purely repulsive $U_{ns}$. There is still an entropic barrier (green line) of about 10$k_BT$ between two fixed points of the $F(z)$.

Not surprisingly, the absence of the SPB transition results in qualitatively different dynamic behaviors of the nanoparticle with the confinement size. The lateral diffusion of the nanoparticle gradually slows down as $H$ decreases with the gradual decrease in $D_\parallel/D_s$ (Fig.~\ref{fig:diffusion}C), consistent with previous studies~\cite{barrat1999large}. As also expected, there is no dramatic change in the solvation state of the purely repulsive nanoparticle, only residing near the walls, with $N_s$ (grey squares in Fig.~\ref{fig:diffusion}A) changing gradually between 35 and 50.

Figures~\ref{fig:interaction}C and D further show how the solvent-solvent interaction $U_{ss}$ with varying $\epsilon_{ss}$ modulates the SPB transition ($\epsilon_{ns}=1k_BT$ and $r_c=2.5\sigma_{ns}$ are fixed). The increasing $\epsilon_{ss}$ pushes the nanoparticle to the walls from the center: $z^*=0$ is globally stable for small $\epsilon_{ss}$ but becomes metastable for large $\epsilon_{ss}$ at both $H=10\sigma_s$ and $8\sigma_s$. Thus, one may control the spatial distribution of the nanoparticle at fixed confinement by changing the \textit{solvent quality}. Such modulation is similar to a gas-liquid phase transition driven by the temperature since $\epsilon_{ss}$ plays the role of an effective temperature. Overall, Fig.~\ref{fig:interaction} suggests that the SPB transition is determined by the delicate balance between the interparticle interactions: when $U_{ss}$ dominates over $U_{ns}$, the nanoparticle prefers to bind to the walls, while it fluctuates at the confinement center when $U_{ns}$ dominates over $U_{ss}$. Our additional calculations further support our argument, where the solvent density can drive the SPB transition at fixed $H$; see the Supporting information.

Finally, we investigate the effects of the confinement interface on the SPB transition (Fig.~\ref{fig:wall}), including the wall structures and the wall interactions $U_{nw}$. Fig.~\ref{fig:wall}A shows that the SPB transition, according to the $F(z)$, occurs regardless of the interface structures (smooth, hexagonally, and tetragonally corrugated surfaces): the globally stable fixed points change from the confinement center to the interface as $H$ decreases. In the case of the smooth walls, the equilibrium states in $F(z)$ shift closer to the interface, implying the difference in the $U_{nw}$ leads to different $\alpha$ constants and thereby fixed points of Eq.~\ref{eq:spb}.

Figure~\ref{fig:wall}B displays the effect of $U_{nw}$ with an asymmetry in the confinement. The slit still consists of the flat walls but with one attractive wall ($r_c=2.5\sigma_{nw}$) and the other purely repulsive wall ($r_c=1.122\sigma_{nw}$) to the nanoparticle. Both walls are yet purely repulsive to the solvent particles with $r_c=1.122\sigma_{sw}$. The attractive $U_{nw}$ can be considered as an additional external force acting on the nanoparticle; such a force, breaking the mirror symmetry, can be expressed as even-order terms in Eq.~\ref{eq:spb}. Not surprisingly, the attractive $U_{nw}$ makes the nanoparticle more likely to stay near the attractive wall: the fixed point near the attractive wall becomes more stable with a lower value of $F(z)$ than near the repulsive wall. Nevertheless, the SPB transition still occurs with such an asymmetry: when $H>H_c$, the globally stable fixed point is still $z^*=0$, and the fixed points near the interfaces become stable only when $H<H_c$.

In summary, from a systematic investigation using molecular dynamics simulations, we show that the spatial distribution of the confined nanoparticle with explicit solvent particles in slit confinement undergoes the first-order phase transition well described by the SPB transition. Across the critical slit size, the discontinuous dynamic transition also occurs, accompanying the change in the solvation state of the nanoparticle. The discontinuous changes in static and dynamic properties of the confined nanoparticle call for the development of molecular hydrodynamics to small nanoparticle solutions under strong confinement, such as the solvent coarse-graining models~\cite{klippenstein2022cross}.
It is also worthwhile for future studies investigating the effects of other external conditions on the physicochemical properties of small nanoparticle solutions in strong confinement, such as finite concentrations of nanoparticles and mechanical driving forces~\cite{koch2024nonequilibrium}.

We thank Taejin Kwon for carefully reading the manuscript. This work was supported by the National Research Foundation of Korea(NRF) grant funded by the Korea government(MSIT) (RS-2024-00338551). This research was supported by the Basic Science Research Program through the National Research Foundation of Korea (NRF) funded by the Ministry of Education (2018R1A6A1A0 3024940). This work was supported by Korea Environment Industry \& Technology Institute(KEITI) through Advanced Technology Development Project for Predicting and Preventing Chemical Accidents Program, funded by Korea Ministry of Environment(MOE)(RS-2023-00219144). J.K. acknowledges the support of the
Korea Institute of Science and Technology Information
(KISTI) Supercomputing Center (KSC-2024-CRE-0168) for the computation.


\begin{suppinfo}\label{SI}

Supporting information includes potentials of mean force with their enthalpic and entropic contributions, the convergence of the time-dependent lateral diffusion coefficients in the long-time limit, and the effect of solvent density on the SPB transition

\end{suppinfo}

%


\end{document}